\documentstyle[12pt,epsfig]{article}
\def\lbldef#1#2{\expandafter\gdef\csname #1\endcsname {#2}}

\def\href#1#2{#2}  

\begin{document}
\baselineskip=15.5pt
\pagestyle{plain}
\setcounter{page}{1}
\begin{titlepage}

\vspace*{-3cm} \date{February, 1999} \hfill 
\raggedleft{CERN-TH/99-48 \\ hep-th/9902208}

\begin{center}

\vskip 5cm

{\Large {\bf Partially Localized Intersecting BPS Branes}}

\vskip 1.5cm

{\large Donam Youm
\footnote{email: Donam.Youm@cern.ch}}

\vskip 0.5cm

Theory Division, CERN,
CH-1211, Geneva 23, Switzerland 

\vskip 2cm

\begin{abstract}
We present the explicit forms of supergravity solutions for 
the various intersecting two BPS branes in eleven and ten 
dimensions, where one brane is localized at the delocalized other  
brane.  Our partially localized supergravity solutions describe brane 
configurations in the near horizon region of the delocalized branes, 
where the two constituent branes meet in the overall transverse space 
and the delocalized branes coincide.  We also give the brane worldvolume 
interpretations for some of such supergravity solutions.  
\end{abstract}
\end{center}
\end{titlepage}

\newpage
\def\ads{{\it AdS}}
\def\adsp{{\it AdS}$_{p+2}$}
\def\cft{{\it CFT}}

\newcommand{\beq}{\begin{equation}}
\newcommand{\eeq}{\end{equation}}
\newcommand{\ber}{\begin{eqnarray}}
\newcommand{\cN}{{\cal N}}
\newcommand{\eer}{\end{eqnarray}}
\newcommand{\cD}{{\Delta}}

\section{Introduction}

Over the past years, there has been active development in 
constructing classical supergravity solutions of $p$-branes
and other solitons in string theories.  Such classical
solutions in string theories made it possible to study
quantum aspects of black holes such as statistical
interpretation of black hole entropy and absorption and
decay rates of black holes within the framework of
string theories.  In such studies, black holes are
regarded as being obtained by wrapping (intersecting)
higher-dimensional $p$-branes around compact manifolds. 
In the course of compactifying $p$-branes, the supergravity
solutions become delocalized along the compactified directions
\footnote{When compactifying the supergravity solutions, one 
has to take an infinite uniform array of branes along the 
directions to be compactified in order to have an isometry  
necessary for the compactification.  Also, by definition, 
compactification means the identification of the corresponding 
points in each cell of the compactification lattice, thereby 
the branes are periodically distributed along the compactified 
directions.  Therefore, in the small size limit of the 
compactification manifold, the supergravity solutions become 
delocalized along the compactified directions.}, 
which include relative transverse directions (the transverse directions 
which are longitudinal to some of other constituent branes)
and possibly some of overall transverse directions. 
So, the corresponding intersecting $p$-brane solutions in
higher dimensions become localized only along the overall transverse 
directions.  The most of intersecting brane solutions that have been 
constructed are such delocalized type.  For the purpose of studying 
black hole physics in string theories, mostly it has been therefore 
sufficient to consider delocalized intersecting brane solutions. 

In the recent AdS/CFT correspondence conjecture \cite{mal} and its 
generalizations (for example \cite{mal2,wit1,wit2}), supergravity 
solutions in string theories also play important roles.  In this conjecture, 
the decoupling limits of the worldvolume theories of brane configurations 
are dual to the supergravity or superstring theories on the manifolds 
of the near horizon geometry of the corresponding supergravity brane 
solutions.  Thereby, one can understand gauge theories, which are the 
decoupling limits (where massive string modes, the Kaluza-Klein (KK) 
modes and gravity modes decouple from the massless modes of open 
strings) of the worldvolume theories of D-branes and other solitons, 
in terms of the supergravity or superstring theories on the near 
horizon manifolds, and vice versa.  In the $D$-brane interpretation 
of gauge theories, the transverse locations of $D$-branes are 
interpreted as moduli of the gauge theories.  For example, transverse 
locations of $N$ numbers of $D$-branes, which are interpreted as scalar 
fields in the $U(N)$ gauge theories, parameterize the Coulomb branch.  
When more than one types of branes intersect, whereas locations of 
the ``light'' $D$-branes are dynamical moduli, locations of the 
``heavy'' branes become couplings, i.e. mass of quarks, since the 
kinetic energy of their excitations is infinite, thereby being 
frozen at their classical values.  

Therefore, if the AdS/CFT correspondence (or generally bulk/boundary 
holographic correspondence) is correct, then all the parameters of the 
gauge theories, which are boundary theories at infinity of the AdS/CFT 
correspondence, have to be mapped one-to-one to the corresponding  
parameters of the corresponding supergravity solutions in the bulk, 
namely the locations of the constituent branes along the transverse 
directions.  However, in delocalized intersecting $p$-brane solutions, 
locations of constituent branes along the worldvolume directions of 
the other branes are not specified, thereby not suitable for studying 
the bulk/boundary correspondence. 

Some attempts have been made to construct localized intersecting 
brane solutions \cite{tsey1,tsey2,yang} with the restricted metric 
Ansatz which has the same form as the corresponding delocalized 
intersecting BPS brane solutions.  Consistency of equations of motion 
along with such simplified metric Ansatz requires that one of the 
branes has to be delocalized on the relative transverse directions.  
Even with such simplified metric Ansatz, however, obtaining the 
``explicit'' analytical form of localized solutions is almost an 
impossible task, since harmonic functions
\footnote{Harmonic function is defined as a solution $f(x_i)$ to the 
Laplace's equation ${1\over{\sqrt{g}}}\partial_i(\sqrt{g}g^{ij}
\partial_j)f(x_i)=0$, where $g\equiv {\rm det}(g_{ij})$.  But in 
the following we will continue to call the solutions to such 
coupled differential equations as harmonic functions.} 
that specify constituent branes now satisfy coupled partial 
differential equations instead of the Laplace's equations in 
the flat (overall) transverse space.  Solutions to such differential 
equations in general do not have simple explicit form in terms of 
elementary functions.  In the case of localized intersecting non-extreme 
brane solutions, even the metric Ansatz (in terms of harmonic functions) 
which would generalize such restricted metric Ansatz for the BPS case is 
not known, not to mention the explicit expressions for harmonic functions.  
``Partially'' localized intersecting brane solutions, where constituent 
branes are localized along the relative transverse directions but delocalized 
along the overall transverse directions, have been constructed 
\cite{khu,gkt,ett,airv}.  However, these solutions satisfy different 
intersecting rules from the ordinary intersecting branes and therefore 
they are only useful for studying brane configurations with exotic 
intersecting rules.  

However, in some special cases in the near-core limit of one 
of constituent branes, one can construct explicit solutions.  
The first attempt was made in Ref. \cite{ity}, where the use was made 
of the fact that the near-core limit or the large charge limit of 
the KK monopole in $D=11$ is $M^{(6,1)}$ times an $A_{N-1}$ 
singularity, which can be obtained from the $D=11$ flat spacetime 
by the $Z_{N}$ identification.  After the dimensional reduction, this 
$A_{N-1}$ singularity metric becomes  near-horizon limit of either 
$D6$-branes or the KK monopole in $D=10$.  Since an $A_{N-1}$ 
singularity is Ricci flat (thereby satisfying the Einstein's equations), 
one can just replace the flat transverse space of $M$-branes by an 
$A_{N-1}$ singularity.  After the compactification down to $D=10$, 
the resulting solutions are localized brane solutions in the core of 
either $D6$-branes or the KK monopoles.  The other type of localized 
intersecting brane solutions that were explicitly constructed is 
the $D(-1)$-brane solution in the core of $D3$-brane  
\cite{chw,bgkr,kl,ps}.  In this case, it was possible to obtain the 
explicit form of the harmonic function for $D(-1)$-brane, since the 
equations of motion reduce to the Laplace's equation in the 
background of the near-horizon geometry of $D3$-branes, i.e. AdS$_{5}$ 
space, which is conformaly flat.  

One may argue that other localized intersecting BPS brane solutions 
(in the core of one of the constituents) can be obtained by just 
applying duality transformations on the above localized solutions.  
However, in order to apply $T$-duality transformations in the 
transverse directions
\footnote{The $T$-duality transformations on supergravity brane 
solutions along the longitudinal directions are not allowed, since 
these transformations introduce additional transverse directions, 
which the resulting transformed supergravity solutions have to 
depend on.  One cannot arbitrarily let the solutions depend on 
this new coordinates through naive generalization of the form of the 
harmonic functions before the $T$-duality transformations, as the 
satisfaction of the field equations is not always guaranteed.  
In fact, the true ``localized'' harmonic functions, which satisfy 
the equations of motion, take different forms from the harmonic 
functions before the $T$-duality transformation on the longitudinal 
directions.  This can be seen from the various intersecting brane 
solutions involving $D$-branes, which are presented in the following.}, 
one has to first compactify this direction, which becomes 
delocalized through smearing or uniform array of branes along this 
direction.  Thereby, the power of radial coordinate in the 
harmonic function (of the above $D6$-brane, KK monopole and the 
$D3$-brane) changes.   This implies that after the required 
delocalization along the $T$-duality direction the near-horizon 
geometries of $D6$-brane and $D=10$ KK monopole do not get uplifted 
to an $A_{N-1}$ singularity and the near-horizon geometry of 
$D3$-branes is no longer conformaly flat.  So, the above tricks for 
constructing localized intersecting branes cannot be applied.  
Therefore, one has to construct such localized intersecting BPS brane 
solutions case by case.  It is the purpose of this paper to construct 
various explicit partially localized intersecting BPS brane solutions 
in the core of the delocalized constituent in various dimensions.  
We will apply simple coordinate transformations to the differential 
equations satisfied by the harmonic functions in order to bring them 
to the forms of partial differential equations which have known explicit 
solutions.  

The paper is organized as follows.  In section 2, we setup the 
general formalism for constructing partially localized BPS intersecting 
brane solutions where one of the constituents is delocalized.  We will 
apply this formalism to construct various explicit partially localized 
intersecting $M$-brane solutions in section 3, and various partially 
localized intersecting brane solutions in ten dimensions in section 
4.  Also, in sections 3 and 4, we discuss some worldvolume 
interpretations of such partially localized intersecting brane 
solutions for the purpose of illustrating possible usefulness of our 
solutions in studying the bulk/boundary holographic correspondence.  
Namely, although partially delocalized, our solutions still contain 
the corresponding parameters of the gauge theories on the ``boundary'', 
which delocalized intersecting brane solutions lack.  In these 
sections, we identify the parameters of our supergravity solutions with 
the corresponding parameters in the gauge theories on the boundary.  

\section{General Setup}

In this section, we setup general formalism, which can be applied 
to any types of intersecting branes in any dimensions, for 
obtaining the explicit expressions for harmonic functions. 

In the BPS case, supergravity solutions for partially localized 
intersecting branes can be obtained with the same metric Ans\"atze and 
the same harmonic superposition rules as the delocalized cases.  
So, in the following, we shall assume the same forms of the metric 
Ans\"atze (which differ for different types of constituent branes and 
therefore will be given in sections 3 and 4 case by case) 
as the delocalized intersecting BPS branes.   Schematically, in 
general the intersecting brane configuration is given by the 
following table.  (This table is given also for the purpose of 
fixing the notations for the spacetime coordinates, which we shall 
follow in the following sections.)  
\begin{center}
\begin{tabular}{|l||c|c|c|c|c|} \hline
{} \ & \ $t$ \ & \ $\vec{w}$ \ & \ $\vec{x}$ \ & \ $\vec{y}$ \ & \ $\vec{z}$ 
\\ \hline\hline
brane 1 \ & \ $\bullet$ \ & \ $\bullet$ \ & \ {} \ & \ $\bullet$ \ & \ {}  
\\ \hline
brane 2 \ & \ $\bullet$ \ & \ $\bullet$ \ & \ $\bullet$ \ & \ {} \ & \ {} 
\\ \hline
\end{tabular}
\end{center}
Here, $t$ is the time coordinate, $\vec{w}$ is the possible overall 
longitudinal coordinate, $\vec{x}=(x_{1},\ldots,x_{p})$ [$\vec{y}=(y_{1},
\ldots,y_{q})$] is the relative transverse coordinate for the brane 1 
[the brane 2], and $\vec{z}=(z_{1},\ldots,z_{r})$ is the overall transverse 
coordinate.  Generally, for any type of intersecting brane 1 (with the 
harmonic function $H_{1}=H_{1}(\vec{x},\vec{z})$) and brane 2 (with the 
harmonic function $H_{2}=H_{2}(\vec{y},\vec{z})$) in any dimensions with 
the above configuration, the harmonic functions satisfy the following 
coupled partial differential equations
\footnote{More precisely, the coordinates $\vec{z}$, $\vec{x}$ and 
$\vec{y}$ are the coordinates in which constituent branes are localized.  
Namely, when some of the coordinates are delocalized due to, for example, 
dimensional reduction, the harmonic functions $H_1$ and $H_2$ still satisfy 
the same coupled differential equations (\ref{laplace}) but just 
do not depend on the delocalized coordinates.  As will be seen in the following, 
in some cases it is necessary to delocalize the configuration along some 
of the overall transverse directions for the purpose of localizing one brane 
to the other.  In such cases, $\vec{z}$ is the coordinate in the part of 
the overall transverse space where the intersecting branes are localized.} 
\cite{tsey1,tsey2,lp,ity,yang}:
\begin{eqnarray}
\partial^{2}_{\vec{z}}H_{1}+H_{2}\partial^{2}_{\vec{x}}H_{1}=0, 
\cr
\partial^{2}_{\vec{z}}H_{2}+H_{1}\partial^{2}_{\vec{y}}H_{2}=0,
\label{laplace}
\end{eqnarray}
along with the constraint
\begin{equation}
\partial_{\vec{x}}H_{1}\partial_{\vec{y}}H_{2}=0.
\label{constrnt}
\end{equation}

The constraint (\ref{constrnt}), i.e. $\partial_{\vec{x}}H_{1}=0$ 
or $\partial_{\vec{y}}H_{2}=0$, implies that either ($i$) the two 
branes are delocalized (localized only along the overall transverse 
directions) or ($ii$) while one brane is completely localized 
the other brane has to be localized along the overall transverse 
directions, only.  So, the supergravity solutions with the above 
mentioned simplified metric Ans\"atze have such limited description 
of the microscope brane configurations on the boundary. 
However, this is not a disadvantageous situation for studying bulk/boundary 
correspondence, since the decoupling limit (where only the massless modes 
of open strings, which describe gauge theories, survive) of the brane 
worldvolume theories requires delocalization of some types of brane 
configurations.  For example, for the configurations where one type of 
branes suspends between other type of branes,  the distance between the 
latter type of branes has to approach zero so that the associated massive 
KK modes on the worldvolume theory of the former branes decouple 
from the massless open string modes.  So, these directions, which 
are the relative transverse directions of the former branes, 
become delocalized
\footnote{We assume here that the directions along which the latter 
type of branes stack are compactified with the distance between the 
stack of the branes being the size of the compactification manifold.}.  

Without loss of generality, for the sake of obtaining harmonic 
functions, we assume that the brane 2 is delocalized:
\begin{equation}
\partial^{2}_{\vec{z}}H_{1}+H_{2}\partial^{2}_{\vec{x}}H_{1}=0, 
\ \ \ 
\partial^2_{\vec{z}}H_2=0.
\label{fnllaplac}
\end{equation}
In general, the harmonic function $H_2$ that satisfies the 
second differential equation in (\ref{fnllaplac}) has the form 
$H_2=1+\sum_i{{Q_i}\over{|\vec{z}-\vec{z}_{0\,i}|^{r-2}}}$, 
where $\vec{z}_{0\,i}$ are locations of the $i$-$th$ brane 2 
with charge $Q_i$.  However, in this paper we will consider 
the case in which the brane 2's coincide at the same location 
in the overall transverse $\vec{z}$-directions.  
This is also required as the decoupling limit of some types 
of brane configurations, as pointed out in the previous 
paragraph. Therefore, we choose the following form of the harmonic 
function $H_2$ in the near core region of the brane 2:
\begin{equation}
H_2={Q\over{z^{r-2}}},
\label{harmtwo}
\end{equation}
where $z\equiv |\vec{z}-\vec{z}_0|$ and $\vec{z}_0$ is the 
location of the stack of $N_2$ brane 2's.  
There might exist solutions for the harmonic function $H_1$ 
in the case where each brane 2's are located at different 
points along the $\vec{z}$-directions.  However, it may not 
be possible to find expression for $H_1$ in closed form in terms 
of elementary functions.  Furthermore, naive inspection of 
the structure of the differential equation satisfied by 
$H_1$ seems to indicate that in general there does not exist a 
closed form of ``localized'' solution for $H_1$ in terms of elementary 
functions where brane 1's are located at arbitrary locations along the 
$\vec{z}$-directions away from the brane 2. 

Then, the first differential equation in (\ref{fnllaplac}) reduces to 
the following form:
\begin{equation}
Q^{-1}z^{-1}\partial_{z}(z^{r-1}\partial_z H_{1})+
\partial^{2}_{\vec{x}}H_{1}=0.
\label{brntwlpl}
\end{equation}
In the case where the harmonic function $H_1$ depends on its 
relative transverse coordinates $\vec{x}$ only through the 
radial coordinate $x\equiv |\vec{x}|$, this differential equation 
becomes of the following form:
\begin{equation}
Q^{-1}z^{-1}\partial_{z}(z^{r-1}\partial_z H_{1})+
x^{-p+1}\partial_{x}(x^{p-1}\partial_xH_{1})=0. 
\label{brntwlplrad}
\end{equation}
We will first obtain solution $H_1=H_1(x,z)$ to the latter differential 
equation (\ref{brntwlplrad}).  This solution can be easily generalized 
as a general solution $H_1=H_1(\vec{x},z)$ to the former differential 
equation (\ref{brntwlpl}).  

The differential equation (\ref{brntwlplrad}) can be solved by transforming 
it to either of the following forms:
\begin{equation} 
[X^{-a}(\partial_XX^a\partial_X)+Y^{-b}(\partial_YY^b\partial_Y)]F(X,Y)=0,
\label{pdeone}
\end{equation}
\begin{equation}
[W^c\partial^2_W+Z^d\partial^2_Z]G(W,Z)=0,
\label{pdetwo}
\end{equation}
where $a$, $b$, $c$ and $d$ are real numbers.  These two partial 
differential equations are related through the following coordinate 
transformations:
\begin{equation}
\left\{\matrix{W=\left({{2-c}\over{2}}X\right)^{2\over{2-c}}\cr 
Z=\left({{2-d}\over{2}}Y\right)^{2\over{2-d}}}\right.,
\ \ \ \ \ \ \ \ 
\left\{\matrix{X=(1-a)W^{1\over{1-a}}\cr 
Y=(1-b)Z^{1\over{1-b}}}\right.,
\label{coordtran}
\end{equation}
where the constants $(a,b)$ and $(c,d)$ are related as
\begin{equation}
a={c\over{c-2}},\ \ \ \ b={d\over{d-2}}.
\label{constrel}
\end{equation}
For a localized brane, the corresponding harmonic function has to be 
of the non-trivial form which is neither sum nor product of a 
function of $X$ (or $W$) and a function of $Y$ (or $Z$)
\footnote{Obtaining the solutions to the partial differential equations 
(\ref{pdeone}) and (\ref{pdetwo}) by applying the method of the additive or 
multiplicative separation of variables is very straightforward.  However, 
when the harmonic functions are of such forms, either the supergravity 
solution does not match onto a delta-function brane source \cite{tsey2} 
or the point singularity of each term in the harmonic function 
represents the brane that is delocalized in the other directions.  The 
possibility of getting solutions with the additive separation of 
variables were pointed out in Ref. \cite{bran}. Also, the expressions for 
harmonic functions in terms of the infinite series of special functions 
by applying the method of multiplicative separation of variables were 
obtained in Refs. \cite{lp,sm}.}.  
Such non-trivial solutions of the partial differential equations 
(\ref{pdeone}) and (\ref{pdetwo}) are respectively given by
\begin{equation}
F(X,Y)=1+{P\over{(X^2+Y^2)^{{a+b}\over 2}}},
\label{solone}
\end{equation}
\begin{equation}
G(W,Z)=1+{P\over{[{4\over{(c-2)^2}}W^{2-c}+{4\over{(d-2)^2}}
Z^{2-d}]^{{cd-c-d}\over{(c-2)(d-2)}}}}, \ \ 
(c\neq 2\neq d),
\label{soltwo}
\end{equation}
where the integration constants in the constant terms have been
set to 1 so that the harmonic functions take ordinary forms.  
One can solve the differential equation (\ref{brntwlplrad}) by transforming 
it to the either of the forms (\ref{pdeone}) and (\ref{pdetwo}).  

The differential equation (\ref{brntwlplrad}) can be put into the 
form (\ref{pdeone}) through the following change of variables:
\begin{equation}
x\to X=x,\ \ \ \  
z\to Y={{2\sqrt{Q}}\over{|4-r|}}z^{{4-r}\over 2},
\label{chvarone}
\end{equation}
resulting in the form (\ref{pdeone}) with $a=p-1$ and $b=r/(4-r)$.  
Or the equation (\ref{brntwlplrad}) can be put into the 
form (\ref{pdetwo}) by applying the following change of variables:
\begin{equation}
x\to W=\left({x\over{2-p}}\right)^{2-p},\ \ \ \ 
z\to Z=\left[{Q\over{(2-r)^2}}\right]^{{2-r}\over{4-r}}
z^{2-r},
\label{chvartwo}
\end{equation}
resulting in the form (\ref{pdetwo}) with $c=(2p-2)/(p-2)$ and 
$d=r/(r-2)$.  Therefore, the harmonic function $H_1=H_1(x,z)$ that 
satisfies the differential equation (\ref{brntwlplrad}) has the form:
\begin{equation}
H_1(x,z)=1+{P\over{[x^2+{{4Q}\over{(4-r)^2}}z^{4-r}]^{{1\over 2}
(p-1+{r\over{4-r}})}}}.
\label{harmh1}
\end{equation}
The first transformation (\ref{chvarone}) already indicates that the 
harmonic function (\ref{harmh1}) for the brane 1 localized at the brane 
2 is not valid when the dimensionality of the the overall transverse 
space is 4 ($r=4$).  Also, the second transformation (\ref{chvartwo}) 
indicates that our method cannot be applied when the overall transverse 
space is two-dimensional ($r=2$), in which case $H_2$ in Eq. (\ref{harmtwo}) 
is logarithmic, and the solution (\ref{harmh1}) is not valid.

Note, as pointed out in the previous paragraph, when the 
dimensionality $r$ of the overall transverse space with the coordinates 
$\vec{z}$ is 4, the coordinate transformations (\ref{chvarone}) and the 
expression (\ref{harmh1}) for the harmonic function $H_1$ are not 
valid.  The only non-trivial solution, which is not a product of a 
function of $x$ and a function of $z$, to the differential equation 
(\ref{brntwlplrad}) with $r=4$ that we have found so far has the 
form:
\begin{equation}
H_1(x,z)=1+P(x^2-pQ\ln z),
\label{singharm}
\end{equation}
although this looks quite extraordinary as a harmonic function 
associated with (localized) branes and this implies that the associated 
brane is delocalized.  However, when the solution is smeared in one of the 
overall transverse directions $\vec{z}$, one can find more acceptable 
expression for the harmonic function $H_1$ from Eq. (\ref{harmh1}) 
(with $r=3$), just as in Ref. \cite{ity}, which represents brane 1's 
that are completely localized except for one delocalized overall 
transverse direction.  

Also, for some other intersecting brane configurations to be discussed 
in the following sections, we note that the corresponding supergravity 
solutions become delocalized since the power in the harmonic function 
$H_1$ is positive instead of negative.  This happens when the overall 
transverse space has large enough dimensionality, i.e. when $r>4$ as 
can be seen from Eq. (\ref{harmh1}).  As can be in seen in the general 
expression for the harmonic function $H_1(x,z)$ in Eq. 
(\ref{harmh1}), which is valid for any type of brane (except for the 
case $r=4$) in any spacetime dimensions, such delocalization depends 
on the dimensionality of the overall transverse space and possibly 
that of the relative transverse space of the brane 1, independently 
of the dimensionality of the overall longitudinal space.  If one 
delocalizes some of the overall transverse directions, the power in the 
harmonic function $H_1$ becomes negative, thereby describing the brane 1 
that are completely localized except for the delocalized overall transverse 
directions.  

This seems to imply that, if there do not exist other class 
of localized solutions (where the brane 1 and the brane 2 meet in the 
overall transverse space), then the corresponding microscopic brane 
configurations have to be delocalized in the relative transverse directions
\footnote{This possibility was later studied in Ref. \cite{mp} by 
noticing such properties of the solutions presented in this paper.} 
unless some of the overall transverse directions are delocalized.  
So, our delocalized intersecting brane solutions (with $r\geq 4$) 
correctly describe the corresponding delocalized microscope brane 
configurations on the boundary.  On the other hand, this might be due 
to our choice of the simplified form of the metric Ansatz, which is the 
same form as the delocalized solutions.  With more general form of metric 
Ansatz, there might exist completely localized intersecting brane solutions 
without any delocalized overall transverse directions.  

In the case when some of the overall transverse directions are delocalized 
for the purpose of localizing the brane 1 at the brane 2, it is understood 
in the following sections that $\vec{z}$ in the harmonic functions is the 
coordinates for the part of the overall transverse directions where the brane 
configuration is localized.  

It can be proven that the original differential equation (\ref{brntwlpl}) 
for the harmonic function $H_1=H_1(\vec{x},\vec{z})$ is solved by
\begin{equation}
H_1(\vec{x},\vec{z})=1+\sum_i{{P_i}\over{[|\vec{x}-\vec{x}_{0\,i}|^2
+{{4Q}\over{(4-r)^2}}|\vec{z}-\vec{z}_{0}|^{4-r}]^{{1\over 2}
(p-1+{r\over{4-r}})}}},
\label{genharm}
\end{equation}
where the $i$-$th$ brane 1 with charge $P_i$ is located at 
$(\vec{x},\vec{z})=(\vec{x}_{0\,i},\vec{z}_{0})$.  Note, this 
general form of the ``modified'' harmonic function for the brane 1 
is for the near horizon region ($|\vec{z}-\vec{z}_0|\approx 0$) of 
the brane 2 when the brane 1 and the brane 2 meet in the overall 
transverse space and when all the brane 2's coincide at 
$\vec{z}_0$.

Note, in the above expressions for harmonic functions $H_1$ and $H_2$, 
$Q$ and $P_i$ are just integration constants that result from solving 
the differential equations.  When one wants to study the partially 
localized intersecting brane solutions in this paper within the frameworks 
of string or M theory, one has to express the constants $Q$ and $P_i$ in 
terms of the numbers $N_1$ and $N_2$ of the brane 1 and brane 2, and the 
charge quantization constants (expressed in terms of the fundamental string 
scale $l_s$, the string coupling constant $g_s$, the eleven dimensional 
Planck scale $l_p$, etc.)  It turns out that although the constant $Q$ of 
$H_2$ is proportional to the number $N_2$ of brane 2's, the constant $P$ of 
$H_1$ is related to the product $N_1N_2$ of the numbers $N_1$ and $N_2$ 
of brane 1's and brane 2's
\footnote{I would like to thank Y. Oz for pointing out this point.}. 

When more than two branes intersect, one can apply the above 
procedure with coupled differential equations with constraints 
that generalize (\ref{laplace}) and (\ref{constrnt}). (For 
example, see Ref. \cite{lp}.)  For such general cases, one may 
have to transform the coupled partial differential equations 
to one of the following forms:
\begin{equation} 
[\sum_iX^{-a_i}_i(\partial_{X_i}X^{a_i}_i\partial_{X_i})]F(X_i)=0,
\label{gpdeone}
\end{equation}
\begin{equation}
[\sum_iW^{c_i}_i\partial^2_{W_i}]G(W_i)=0,
\label{gpdetwo}
\end{equation}
where $a_i$ and $c_i$ are real numbers. The solutions to these partial 
differential equations are respectively given by
\begin{equation}
F(X_i)=1+P/(\sum_iX^2_i)^{{\sum_ia_i}\over 2},
\label{gsolone}
\end{equation}
\begin{equation}
G(W_i)=1+P/[\sum_i{4\over{(c_i-2)^2}}
W^{2-c_i}_i]^{\sum_i{{c_i}\over{2(c_i-2)}}}, \ \ 
(c_i\neq 2).
\label{gsoltwo}
\end{equation}

In the following sections, we will present the explicit forms of 
every possible partially localized intersecting two brane solutions 
in eleven and ten dimensions in the case when the explicit form of 
the ``modified'' harmonic function can be obtained by applying the 
method discussed in this section.  Localized intersecting brane 
solutions in other dimensions can be similarly constructed just 
by applying the procedure discussed in this section.  All the 
possible delocalized intersecting brane configurations are studied in 
Refs. \cite{bran1,bran2}, which we generalize to the localized case.  
In the following, we will just write down expressions for spacetime 
metric and the explicit forms of harmonic functions, since the 
expressions for other fields (dilaton and form fields) can be 
straightforwardly constructed by applying the same harmonic function 
superposition rules as the delocalized intersecting brane cases but 
now with new ``localized'' harmonic functions replaced.  

\section{Localized Intersecting $M$-branes}

In eleven dimensions, the basic constituents of intersecting branes 
are $M2$- and $M5$-branes, which respectively carry electric 
and magnetic charges of the three-form field, the KK monopole 
and the pp-wave.  In the following, we write down the spacetime 
metrics for all the possible combinations of intersecting pairs 
of these branes along with the explicit forms of the harmonic 
functions. 

\subsection{Intersecting $M2$- and $M5$-branes}

There are 4 overall transverse directions ($r=4$).  The $M2$-branes
have 4 relative transverse directions ($p=4$), and the $M5$-branes 
have 1 relative transverse direction ($q=1$).  Since this configuration 
is interpreted as $M2$-branes ending on $M5$-branes, it is natural 
to let the solution be delocalized along the relative transverse 
direction of the $M5$-branes.  Formally, one can however construct 
solution for the other case, as well.  The spacetime metric has the 
following form:
\begin{eqnarray}
ds^2_{11}&=&H^{1/3}_2H^{2/3}_5[(H_2H_5)^{-1}(-dt^2+dw^2)+
H^{-1}_5(dx^2_1+\ldots+dx^2_4)
\cr
& &+H^{-1}_2dy^2+dz^2_1+\ldots+dz^2_4],
\label{m2m5}
\end{eqnarray}
where the harmonic functions $H_2$ and $H_5$ are respectively associated 
with the $M2$- and $M5$-branes.
For the purpose of obtaining more physically acceptable form of 
solution, we delocalize the solution along one of the overall transverse 
directions $\vec{z}$
\footnote{When the solution is completely localized along all the 
overall transverse directions, the harmonic function takes an 
unacceptable form (\ref{singharm}) with a logarithm}.  
Then, the harmonic functions are given by
\begin{equation}
H_2=1+\sum_i{{Q_i}\over{(|\vec{x}-\vec{x}_{0\,i}|^2+4P
|\vec{z}-\vec{z}_{0}|)^3}},\ \ \ \ 
H_5={P\over{|\vec{z}-\vec{z}_{0}|}}.
\label{m2m5harm}
\end{equation}

The effective worldvolume theory of the $M5$-brane is the $5+1$ 
dimensional $(2,0)$ tensor multiplet containing 5 scalars, an 
anti-symmetric tensor with self-dual 3-form field strength, and 4 
chiral fermions.  The ends of $M2$-branes on the $M5$-branes are 
regarded as (self-dual) strings (in the $M5$-brane worldvolume 
theory) charged under this tensor. The charges $Q_i$ and locations 
$\vec{x}_{0\,i}$ in the above supergravity solution are related 
to charges and locations of these strings in the worldvolume theory.

\subsection{Intersecting two $M5$-branes}

There are 3 overall transverse directions ($r=3$) and 
2 relative transverse directions for each $M5$-brane ($p=2=q$).  
So, in the core region of one of the $M5$-branes, the metric 
has the following form:
\begin{eqnarray}
ds^2_{11}&=&(H_1H_2)^{2/3}[(H_1H_2)^{-1}(-dt^2+dw^2_1+dw^2_2+dw^2_3)
\cr
& &+H^{-1}_2(dx^2_1+dx^2_2)+H^{-1}_1(dy^2_1+dy^2_2)+dz^2_1+dz^2_2+dz^2_3],
\label{m5m5}
\end{eqnarray}
with the harmonic functions given by:
\begin{equation}
H_1=1+\sum_i{{P_i}\over{[|\vec{x}-\vec{x}_{0\,i}|^2+
4P|\vec{z}-\vec{z}_{0}|]^2}}, \ \ \  
H_2={P\over {|\vec{z}-\vec{z}_{0}|}}.
\label{m5m5harm}
\end{equation}

From the perspective of $M5$-brane worldvolume theory, the 
3-dimensional intersection is $1/2$ supersymmetric 3-branes 
that carry the self-dual 3-form central charges of the 
worldvolume superalgebra of the $M5$-brane \cite{bgt}.  
The parameters $P_i$ and $\vec{x}_{0\,i}$ are related to the 
charges and the locations of these 3-branes in the $M5$-brane 
worldvolume theory. 

\subsection{Intersecting two $M2$-branes}

In this case, there are 6 overall transverse directions ($r=6$) 
and the dimensions of the relative transverse spaces of both 
of $M2$-branes are 2 ($p=2=q$).  In the core region of one of 
the $M2$-branes, the solution has the following form:
\begin{eqnarray}
ds^2_{11}&=&H^{1/3}_1H^{1/3}_2[-H^{-1}_1H^{-1}_2dt^2+
H^{-1}_2(dx^2_1+dx^2_2)+H^{-1}_1(dy^2_1+dy^2_2)
\cr
& &\ \ \ \ \ +dz^2_1+\ldots+dz^2_6],
\label{m2m2}
\end{eqnarray}
where the harmonic functions are given by
\begin{equation}
H_1=1+\sum_iQ_i[|\vec{x}-\vec{x}_{0\,i}|^2+Q|\vec{z}-\vec{z}_{0}|^{-2}], 
\ \ \ \ 
H_2={Q\over{|\vec{z}-\vec{z}_{0}|^4}}. 
\label{m2m2harm}
\end{equation}
When $n$ of the overall transverse directions are delocalized, 
the harmonic functions take the following form:
\begin{equation}
H_1=1+\sum_i{{Q_i}\over{[|\vec{x}-\vec{x}_{0\,i}|^2+{{4Q}\over
{(n-2)^2}}|\vec{z}-\vec{z}_{0}|^{n-2}]^{2\over{n-2}}}},\ \ \ \ 
H_2={Q\over{|\vec{z}-\vec{z}_{0}|^{4-n}}}.
\label{m2m2delocal}
\end{equation}
So, when 3 overall transverse directions are delocalized, 
the harmonic function $H_1$, as well as $H_2$, takes a 
standard form where branes are localized except for the 3 
delocalized overall transverse directions.  

From the perspective of the $M2$-brane worldvolume theory, 
the $0+1$ dimensional intersection is interpreted as a 
0-brane coupled to the zero form central charge in the 
worldvolume superalgebra.  Such 0-brane is charged with respect 
to the Hodge-dual of a transverse scalar.  The parameters 
$\vec{x}_{0\,i}$ and $Q_i$ are related to the locations and 
charges of these worldvolume 0-branes.  

\subsection{$M2$-brane with the KK monopole}

When the flat transverse space of the $M2$-brane is replaced 
by the Taub-NUT terms of the KK monopole, the spacetime metric 
takes the following form:
\begin{eqnarray}
ds^2_{11}&=&H^{-2/3}_2[-dt^2+dw^2_1+dw^2_2]+H^{1/3}_2[dx^2_1+\ldots+
dx^2_4
\cr
& &+H_K(dz^2_1+dz^2_2+dz^2_3)+H^{-1}_K(dy+A_idz_i)^2],
\label{m2kk}
\end{eqnarray}
where the harmonic functions $H_2$ and $H_K$ for the $M2$-brane and 
the KK monopole and a 1-form potential ${\bf A}=(A_i)$ satisfy the 
equations:
\begin{equation}
\partial^{2}_{\vec{z}}H_{K}=0,\ \ \ 
\partial_{z_{i}}H_{K}=\epsilon_{ijk}\partial_{z_{j}}A_{k},\ \ \ 
\partial^{2}_{\vec{z}}H_{2}+H_{K}\partial^{2}_{\vec{x}}H_{2}=0.
\label{m2kklap}
\end{equation}
In the core region of the KK monopole or in the limit of large KK 
monopole charge, the harmonic functions are therefore given by:
\begin{eqnarray}
H_{K}&=&{{Q_{KK}}\over{|\vec{z}-\vec{z}_{0}|}},\ \ \ \ 
{\bf A}=Q_{KK}\cos\theta d\phi,
\cr
H_2&=&1+\sum_i{{Q_i}\over{(|\vec{x}-\vec{x}_{0\,i}|^{2}
+4Q_{{KK}}|\vec{z}-\vec{z}_{0}|)^3}}.
\label{m2kkharm}
\end{eqnarray}
This solution reproduces the one in Ref. \cite{ity}.

The 0-form central charges of the $M2$-brane worldvolume superalgebra 
are equivalent to the space components of the dual 2-forms, which 
are carried by a worldvolume 2-branes.  This 2-brane is interpreted as 
the intersection of the $M2$-brane with the KK monopole \cite{bgt}.  
The locations and the charges of these worldvolume 2-branes are 
related to $\vec{x}_{0\,i}$ and $Q_i$ of the above supergravity 
solutions.

\subsection{$M5$-brane with the KK monopole}

When the flat transverse space of the $M5$-brane is replaced 
by the Taub-NUT terms of the KK monopole, the spacetime metric 
takes the following form:
\begin{eqnarray}
ds^2_{11}&=&H^{-1/3}_5[-dt^2+dw^2_1+\ldots+dw^2_5]
+H^{2/3}_5[dx^2
\cr
& &+H_K(dz^2_1+dz^2_2+dz^2_3)+H^{-1}_K(dy+A_idz_i)^2],
\label{m5kk}
\end{eqnarray}
where the harmonic functions $H_5$ and $H_K$ for the $M5$-branes and 
the KK monopole and a 1-form potential ${\bf A}=(A_i)$ satisfy the 
equations:
\begin{equation}
\partial^{2}_{\vec{z}}H_{K}=0,\ \ \ 
\partial_{z_{i}}H_{K}=\epsilon_{ijk}\partial_{z_{j}}A_{k},\ \ \ 
\partial^{2}_{\vec{z}}H_{5}+H_{K}\partial^{2}_xH_{5}=0.
\label{m5kklap}
\end{equation}
In the core region of the KK monopole or in the limit of large KK 
monopole charge, the harmonic functions are therefore given by:
\begin{eqnarray}
H_{K}&=&{{Q_{KK}}\over{|\vec{z}-\vec{z}_{0}|}},\ \ \ \ 
{\bf A}=Q_{KK}\cos\theta d\phi,
\cr
H_5&=&1+\sum_i{{P_i}\over{(|\vec{x}-\vec{x}_{0\,i}|^{2}+4Q_{{KK}}
|\vec{z}-\vec{z}_{0}|)^{{3/2}}}}.
\label{m5kkharm}
\end{eqnarray}
This solution is also constructed in Ref. \cite{ity}.

The time component of the 1-form central charge in the $M5$-brane 
worldvolume superalgebra can be viewed as the space components of 
a worldvolume dual 5-form.  This charge is carried by the KK monopole 
\cite{bgt}.  Therefore, this configuration is interpreted as the 
$M5$-brane inside of the KK monopole \cite{kk1,kk2}.  

\subsection{The pp wave in the background of $M2$-branes}

For the pp wave which travels along one of the worldvolume 
directions, which we choose to be $w$, of the $M2$-brane, 
the metric takes the following form:
\begin{eqnarray}
ds^{2}_{11}&=&H^{-2/3}_{2}[-dt^{2}+dw^{2}+dx^{2}+
(H_{W}-1)(dt-dw)^{2}]
\cr
& &\ \ \ \ \ +H^{1/3}_{2}(dz^{2}_{1}+\ldots+dz^{2}_{8}),
\label{m2pp}
\end{eqnarray}
with the harmonic functions $H_{2}$ and $H_{W}$ for the $M2$-brane 
and the pp wave satisfying the following equations:
\begin{equation}
\partial^{2}_{\vec{z}}H_{2}=0,\ \ \ \  
\partial^{2}_{\vec{z}}H_{W}+H_{2}\partial^{2}_xH_{W}=0.
\label{m2pplap}
\end{equation}
In the core region of the $M2$-brane, the harmonic functions 
are therefore given by:
\begin{equation}
H_{2}={{Q}\over{|\vec{z}-\vec{z}_{0}|^{6}}},\ \ \ \ 
H_{W}=1+Q_{W}(x^{2}+{Q\over 4}{1\over {|\vec{z}-\vec{z}_{0}|^{4}}}).
\label{m2ppharm}
\end{equation}
When $n$ of the overall transverse directions are delocalized, 
the harmonic functions take the following forms:
\begin{equation}
H_2={Q\over{|\vec{z}-\vec{z}_{0}|^{6-n}}},\ \ \ \ 
H_W=1+{{Q_W}\over{[x^2+{{4Q}\over{(n-4)^2}}
|\vec{z}-\vec{z}_{0}|^{n-4}]^{{8-n}\over{2(n-4)}}}}.
\label{m2ppdelocal}
\end{equation}
So, with 5 of the overall transverse directions delocalized 
the solutions become localized except for these delocalized 
overall transverse directions. 

Upon dimensional reduction to $D=10$, the $D=11$ pp-wave becomes 
$D0$-brane in the type-IIA theory.  The 0-form central charges 
in the $D=11$ pp-wave superalgebra are decomposed into two sets of 
the 0-form central charges in the $D=10$ $D0$-brane superalgebra, 
which can be respectively interpreted in the transverse 9-space 
as the central charges carried by a $D4$-brane or the KK monopole 
and a fundamental string or a $D8$-branes \cite{bgt}.  When 
uplifted to $D=11$, the resulting configurations are $M2$-brane 
and $M5$-brane and the KK monopole involving the pp-wave, whose 
supergravity solutions are presented in this section and the 
following sections.  
On the other hand, the 3-momentum in the $M2$-brane superalgebra, 
interpreted as a 0-form in the transverse 8-space, is the null 
3-momentum of the pp-wave inside the $M2$-brane \cite{bgt}.
Similarly, the 5-momentum in the $M5$-brane superalgebra, 
interpreted as a 0-form in the transverse 5-space, is the 
null 5-momentum of the $D=11$ pp-wave inside of the $M5$-brane, 
whose supergravity solution is given in the next subsection.  

\subsection{The pp-wave in the background of $M5$-branes}

For the pp wave which travels along one of the worldvolume 
directions, which we choose to be $w$, of the $M5$-brane, 
the metric takes the following form:
\begin{eqnarray}
ds^{2}_{11}&=&H^{-1/3}_{5}[-dt^{2}+dw^{2}+
(H_{W}-1)(dt-dw)^{2}+dx^{2}_{1}+\ldots+dx^{2}_{4}]
\cr
& &\ \ \ \ \ +H^{2/3}_{5}(dz^{2}_{1}+\ldots+dz^{2}_{5}),
\label{m5pp}
\end{eqnarray}
with the harmonic functions $H_{5}$ and $H_{W}$ for the $M5$-brane 
and the pp wave satisfying the following equations:
\begin{equation}
\partial^{2}_{\vec{z}}H_{5}=0,\ \ \ \  
\partial^{2}_{\vec{z}}H_{W}+H_{5}\partial^{2}_{\vec{x}}H_{W}=0.
\label{m5pplap}
\end{equation}
In the core region of the $M5$-brane, the harmonic functions 
are therefore given by:
\begin{equation}
H_{5}={{P}\over{|\vec{z}-\vec{z}_{0}|^{3}}},\ \ \ \ 
H_{W}=1+Q_{W}(x^{2}+{{4P}\over {|\vec{z}-\vec{z}_{0}|}}).
\label{m5ppharm}
\end{equation}
When $n$ of the overall transverse directions are delocalized, 
the harmonic functions take the following forms:
\begin{equation}
H_5={P\over{|\vec{z}-\vec{z}_{0}|^{3-n}}},\ \ \ \  
H_W=1+{{Q_W}\over{[x^2+{{4P}\over{(n-1)^2}}
|\vec{z}-\vec{z}_{0}|^{n-1}]^{{n+1}\over{n-1}}}}.
\label{m5ppdelocal}
\end{equation}
So, with 2 of the overall transverse directions delocalized, 
the solutions become localized except for these two delocalized 
overall transverse directions.

The pp-wave localized at the $M5$-brane is interpreted from the 
perspective of the worldvolume theory at the boundary as the 
neutral (with respect to the self-dual 3-form field strength in 
the $M5$-brane worldvolume theory) string in the $M5$-brane 
worldvolume.

\subsection{The pp-wave in the background of the KK monopole}

The spacetime metric for the pp wave which propagates in the 
background of the KK monopole has the following form:
\begin{eqnarray}
ds^{2}_{11}&=&-dt^{2}+dw^{2}+(H_{W}-1)(dt-dw)^{2}+
dx^{2}_{1}+\ldots+dx^{2}_{5}
\cr
& &+H_{K}(dz^{2}_{1}+dz^{2}_{2}+dz^{2}_{3})+
H^{-1}_{K}(dy+A_{i}dz_{i})^{2}, 
\label{kkpp}
\end{eqnarray}
where the harmonic functions $H_{K}$ and $H_{W}$ for the 
KK monopole and the pp wave and a 1-form potential 
${\bf A}=(A_{i})$ satisfy the equations:
\begin{equation}
\partial^{2}_{\vec{z}}H_{K}=0,\ \ \ 
\partial_{z_{i}}H_{K}=\epsilon_{ijk}\partial_{z_{j}}A_{k},\ \ \ 
\partial^{2}_{\vec{z}}H_{W}+H_{K}\partial^{2}_{\vec{x}}H_{W}=0.
\label{kkpplap}
\end{equation}
In the core region of the KK monopole or in the limit of large KK 
monopole charge, the harmonic functions are therefore given by:
\begin{eqnarray}
H_{K}&=&{{Q_{KK}}\over{|\vec{z}-\vec{z}_{0}|}},\ \ \ \ 
{\bf A}=Q_{KK}\cos\theta d\phi,
\cr
H_{W}&=&1+{{Q_{W}}\over{(x^{2}+4Q_{{KK}}
|\vec{z}-\vec{z}_{0}|)^{{7/2}}}}.
\label{kkppharm}
\end{eqnarray}

\section{Intersecting Branes in Ten Dimensions}

In ten dimensions, the basic constituents of intersecting branes 
are $D$-branes, fundamental string, solitonic $NS5$-brane, the 
KK monopole and the pp wave.  In the following, we present intersecting 
brane configurations of all the possible combinations.  

\subsection{Two $Dp$-branes self-intersecting over $(p-2)$ dimensions}

In this case, the overall transverse space has $7-p$ dimensions and 
there are 2 relative transverse directions for both of $Dp$-branes. 
The spacetime metric has the following form:
\begin{eqnarray}
ds^2_{10}&=&(H_1H_2)^{-1/2}(-dt^2+dw^2_1+\ldots+dw^2_{p-2})+
H^{1/2}_1H^{-1/2}_2(dx^2_1+dx^2_2)
\cr
& &+H^{-1/2}_1H^{1/2}_2(dy^2_1+dy^2_2)+(H_1H_2)^{1/2}
(dz^2_1+\ldots+dz^2_{7-p}),
\label{dpdp}
\end{eqnarray}
where the harmonic functions $H_1$ and $H_2$ for each $Dp$-branes 
satisfy the following equations:
\begin{equation}
\partial^{2}_{\vec{z}}H_{1}+H_{2}\partial^{2}_{\vec{x}}H_{1}=0, 
\ \ \ 
\partial^2_{\vec{z}}H_2=0.
\label{dpdplp}
\end{equation}
The harmonic functions are therefore given by
\begin{equation}
H_1=1+\sum_i{{Q_i}\over{[|\vec{x}-\vec{x}_{0\,i}|^2+{{4Q}\over
{(p-3)^2}}|\vec{z}-\vec{z}_0|^{p-3}]^{2\over{p-3}}}},\ \ \ 
H_2={Q\over{|\vec{z}-\vec{z}_0|^{5-p}}}.
\label{dpdpharm}
\end{equation}

Note, for $p=2$, this solution becomes delocalized.  So, one has to 
delocalize some of the overall transverse directions to obtain supergravity 
solution representing the intersecting brane localized except for 
the delocalized overall transverse directions.  When the overall transverse 
space is 4-dimensional, i.e. self-intersecting $D3$-branes, the above 
expression for the harmonic function $H_1$ is singular.  In this case, one 
has to delocalize one of the overall transverse directions. 
With two [one] overall transverse directions are delocalized for $p=2$ 
[for $p=3$], the harmonic functions are given by:
\begin{equation}
H_1=1+\sum_i{{Q_i}\over{[|\vec{x}-\vec{x}_{0\,i}|^2
+4Q|\vec{z}-\vec{z}_0|]^2}},\ \ \ 
H_2={Q\over {|\vec{z}-\vec{z}_{0}|}}.
\label{d3d3harm}
\end{equation}

The $Dp$-brane worldvolume theory contains a scalar (interpreted as 
a Goldstone mode of spontaneously broken translational invariance by 
the $Dp$-brane), which is Hodge-dualized to a worldvolume $(p-1)$-form 
potential that the $(p-2)$-brane (common intersection of the two 
$Dp$-branes) couples to.

\subsection{$Dp$-branes ending on $D(p+2)$-branes}

There are $6-p$ overall transverse directions.  The dimensions of 
the relative transverse space are 3 for $Dp$-branes and 1 for 
$D(p+2)$-branes.  Since $Dp$-branes stretch between $D(p+2)$-branes, 
it is natural to let $D(p+2)$-branes to be delocalized along their 
relative transverse direction.  The spacetime metric is given by
\begin{eqnarray}
ds^2_{10}&=&(H_pH_{p+2})^{-1/2}(-dt^2+dw^2_1+\ldots+dw^2_{p-1})
+H^{1/2}_pH^{-1/2}_{p+2}(dx^2_1+dx^2_2+dx^2_3)
\cr
& &+H^{-1/2}_pH^{1/2}_{p+2}dy^2+(H_pH_{p+2})^{1/2}(dz^2_1+\ldots+dz^2_{6-p}),
\label{dpdq}
\end{eqnarray}
where the harmonic functions $H_p$ and $H_{p+2}$ for $Dp$- and 
$D(p+2)$-branes satisfy
\begin{equation}
\partial^{2}_{\vec{z}}H_p+H_{p+2}\partial^{2}_{\vec{x}}H_p=0, 
\ \ \ 
\partial^2_{\vec{z}}H_{p+2}=0.
\label{dpdqlap}
\end{equation}
The harmonic functions are given by
\begin{equation}
H_p=1+\sum_i{{Q_i}\over{[|\vec{x}-\vec{x}_{0\,i}|^2+
{{4Q}\over{(p-2)^2}}|\vec{z}-\vec{z}_{0}|^{p-2}]^{(p+2)\over 
{2(p-2)}}}},\ \ \ H_{p+2}={Q\over {|\vec{z}-\vec{z}_{0}|^{4-p}}}.
\label{dpdqharm}
\end{equation}

The fluctuations of locations $\vec{x}_{0\,i}$ of the $Dp$-branes 
along the $D(p+2)$-brane directions, together with the $y$-component 
of the $Dp$-worldvolume gauge field, forms a massless hypermultiplet 
with free boundary conditions at $y=0$.  The scalars describing 
the fluctuations of location $\vec{z}_0$ of the $Dp$-branes along 
the directions perpendicular to the $D(p+4)$-branes, together with 
the $t$- and $\vec{w}$-components of the $Dp$-worldvolume gauge 
field, form a vectormultiplet with each component field satisfying 
the Dirichlet boundary conditions.  

From the point of view of the $D(p+2)$-branes, the ends of the 
$Dp$-branes are charged objects in the $D(p+2)$-brane worldvolume.  
For $D1$-branes ending on $D3$-branes ($p=1$ case), the ends of 
the $D1$-branes are magnetic sources for the $D3$-brane 
worldvolume gauge fields.  Therefore, the worldvolume theory is 
the magnetic monopoles in the 4-dimensional $U(N_{p+2})$ Yang-Mills 
theory.  The locations $\vec{x}=\vec{x}_{0\,i}$ of the $D1$-branes 
along the directions of the $D3$-branes, together 
with the Wilson lines of the $\sum_iN_{p\,i}$ $U(1)$ worldvolume 
gauge fields of $D1$-branes along the $y$-direction, parameterize  
the monopole moduli space.  The charges $Q_i$ of the $D1$-branes 
are interpreted as magnetic charges of the monopoles.  Under the 
$S$-duality of the type-IIB theory, this supergravity brane 
solution transforms to the supergravity solution for fundamental 
strings ending on $D3$-branes, which is presented in the following 
subsection.  From the $D3$-brane worldvolume point of view, this 
$S$-duality is the Montonen-Olive's strong-weak coupling 
electric-magnetic duality of the $N=4$ super-Yang-Mills theory 
in 4 dimensions.  The ends of the fundamental strings on the 
$D3$-branes are electric charge sources, interpreted as charged 
gauge bosons. Just as the $D1$-brane and the fundamental string 
form a multiplet under the type-IIB $S$-duality, the charged gauge 
boson and the magnetic monopole transform as a multiplet under the 
Montonen-Olive duality.

When $p=1$, the above harmonic function $H_p$ represents delocalized 
$Dp$-branes, thereby requiring delocalization of some of overall 
transverse directions.  When $p=2$, the overall transverse space is 
4-dimensional. So, one of the overall transverse directions has to be 
delocalized.  With two [one] of the overall transverse directions 
delocalized for the $p=1$ [$p=2$] case, harmonic functions are given by:
\begin{equation}
H_p=1+\sum_i{{Q_i}\over{[|\vec{x}-\vec{x}_{0\,i}|^2+
4Q|\vec{z}-\vec{z}_{0}|]^{5\over 2}}},\ \ \ 
H_{p+2}={Q\over {|\vec{z}-\vec{z}_{0}|}}.
\label{d4dqharm}
\end{equation}

\subsection{$Dp$-branes inside of the worldvolume of $D(p+4)$-branes}

There are $5-p$ overall transverse directions.  The dimensionalities 
of the relative transverse spaces are 4 for $Dp$-branes and 
0 for $D(p+4)$-branes.  The spacetime metric has the following form:
\begin{eqnarray}
ds^2_{10}&=&(H_pH_{p+4})^{-1/2}(-dt^2+dw^2_1+\ldots+dw^2_p)+
H^{1/2}_pH^{-1/2}_{p+4}(dx^2_1+\ldots+dx^2_4)
\cr
& &\ \ \ \ +(H_pH_{p+4})^{1/2}(dz^2_1+\ldots+dz^2_{5-p}),
\label{dinsd}
\end{eqnarray}
where the harmonic functions $H_p$ and $H_{p+4}$ for $Dp$- and 
$D(p+4)$-branes satisfy the differential equations:
\begin{equation}
\partial^{2}_{\vec{z}}H_p+H_{p+4}\partial^{2}_{\vec{x}}H_p=0, 
\ \ \ 
\partial^2_{\vec{z}}H_{p+4}=0.
\label{dinsdlap}
\end{equation}
The harmonic functions are therefore given by
\begin{equation}
H_p=1+\sum_i{{Q_i}\over{[|\vec{x}-\vec{x}_{0\,i}|^2+
{{4Q}\over{(p-1)^2}}|\vec{z}-\vec{z}_{0}|^{p-1}]^{{p+1}
\over{p-1}}}},\ \ \ H_{p+4}={Q\over {|\vec{z}-\vec{z}_{0}|^{3-p}}}.
\label{dinsdharm}
\end{equation}

The decoupling limit of the worldvolume theory of the corresponding 
microscopic $D$-brane configuration is ($i$) the $(p+1)$-dimensional 
$U(\sum_{i}N_{p\,i})$ gauge theory with $N_{p+4}$ flavors (in the fundamental 
representation of $U(\sum_{i}N_{p\,i})$) with the $U(N_{p+4})$ gauge 
symmetry of the $D(p+4)$-branes being a global symmetry from the point of 
view of $Dp$-branes, or ($ii$) small $U(N_{p+4})$ instantons from the point 
of view of $D(p+4)$-branes with $\sum_{i}N_{p\,i}$ $Dp$-branes being 
pointlike defects in the fundamental of $U(N_{p+4})$.  

From the point of view of the worldvolume theory of the $Dp$-branes, 
the locations of the branes are interpreted as follows.  The locations 
of the $D(p+4)$-branes relative to the locations of $Dp$-branes in 
the $\vec{z}$-direction are masses $\vec{m}_{j}$ ($j=1,\ldots,N_{p+4}$) for 
the $N_{p+4}$-fundamentals. Since all the $D(p+4)$-branes coincide in the 
above supergravity solution, the full $U(N_{p+4})$ symmetry is left intact, 
while the quarks in the fundamental of $U(N_{p+4})$ remain massless.  
The locations $\vec{x}_{0\,i}$ of $Dp$-branes parallel to the 
$D(p+4)$-branes correspond to expectation values of an adjoint 
hypermultiplet of $U(\sum_{i}N_{p\,i})$.  Since all the $Dp$-branes coincide 
in the direction $\vec{z}$ transverse to the $D(p+4)$-branes at 
$\vec{z}=\vec{z}_{0}$ (parameterizing the Coulomb branch of the 
$U(\sum_{i}N_{p\,i})$ gauge theory) for the above supergravity solution, 
the full $U(\sum_{i}N_{p\,i})$ gauge symmetry is left unbroken.

In the case of $p=3$, i.e. intersecting $D3$- and $D7$-branes, one has 
to include the orientifold 7 plane with 8 units of $D7$-brane charge 
in order to cancel the brane charges.  Then, the $D3$- and $D7$-branes 
have to located in pairs as mirror images of the orientifold plane 
along the $\vec{z}$-directions.  Note, the ground state of an 
open string that stretches between the same $D3$-brane is  
the neutral gauge boson $W^3_{\mu}$, whereas those that stretch 
between a pair of $D3$-brane and its mirror image are the charged 
gauge bosons $W^{\pm}_{\mu}$.  Since all the $D3$-branes in the above 
supergravity solution
\footnote{When $p=3$, the harmonic function $H_7$ for the $D7$-branes 
has to be logarithmic.  Therefore, the solution (\ref{dinsdharm}) 
is not valid and one cannot obtain the explicit expression for 
the harmonic functions applying the method discussed in section 2. 
However, there might exist the explicit expressions for harmonic 
functions where all the $D$-branes coincide in the $\vec{z}$-direction. 
On the other hand, such supergravity solution cannot be trusted 
since the conservation of brane charges requires that there can 
be only 4 $D3$-branes and 4 $D7$-branes.} 
coincide in the $\vec{z}$-direction, these charged gauge bosons are 
massless and therefore the full $SU(2)$ gauge symmetry on the 
worldvolume of $D3$-branes is left unbroken.  Since the relative 
locations of $D7$-branes and their mirror images with respect to 
$D3$-brane locations along the $\vec{z}$-direction is zero, the quarks 
are massless and the $SO(8)$ symmetry on the worldvolume of $D7$-branes 
remains unbroken.  

The $p=2$ case is the supergravity solutions for $D2$-branes within 
$D6$-branes constructed in Ref. \cite{ity}.  
The above solution becomes singular for $p=1$ and delocalized for $p=0$.   
After one [two] of the overall transverse directions are delocalized 
for $p=1$ [for $p=0$], the harmonic functions take the following forms:
\begin{equation}
H_p=1+\sum_i{{Q_i}\over{[|\vec{x}-\vec{x}_{0\,i}|^2+
4Q|\vec{z}-\vec{z}_{0}|]^3}},\ \ \ 
H_{p+4}={Q\over {|\vec{z}-\vec{z}_{0}|}}.
\label{d5inharm}
\end{equation}

One can also obtain the localized BPS solution for $D(-1)$-branes 
in the background of $D3$-branes (the $p=-1$ case) applying the 
the general formalism in section 2, although the harmonic function 
for the $D(-1)$-brane is shown to satisfy the Laplace's equation in 
the AdS$_5$ space.  After the Wick rotation to the Euclidean time 
coordinate $x_0\equiv \tau$, the spacetime metric takes the following 
form:
\begin{equation}
ds^2_{10}=H^{1/2}_{-1}H^{-1/2}_3(dx^2_0+\ldots+dx^2_3)+
(H_{-1}H_3)^{1/2}(dz^2_1+\ldots+dz^2_6),
\label{dinst}
\end{equation}
and the dilaton and the 0-form field in the RR sector are 
respectively given by $e^{\phi}=H_{-1}$ and $(\chi-\chi_{\infty})=
\pm(e^{-\phi}-e^{-\phi_{\infty}})$.  Here, the subscript $\infty$ 
denotes the value of a field at the AdS$_5$ boundary.  
The harmonic functions $H_{-1}$ and $H_3$ for the $D(-1)$-branes and 
$D3$-branes satisfy the following equations:
\begin{equation}
\partial^{2}_{\vec{z}}H_{-1}+H_3\partial^{2}_{\vec{x}}H_{-1}=0, 
\ \ \ 
\partial^2_{\vec{z}}H_3=0,
\label{dinstlap}
\end{equation}
where $\vec{x}=(x_0,\ldots,x_3)$ now the time coordinate $x_0=\tau$ 
included.  This system of partial differential equations cannot 
be solved by applying the method discussed in section 2.  But it 
can be easily proved that the following harmonic functions satisfy 
(\ref{dinstlap}):
\begin{equation}
H_{-1}=1+\sum_i{{Q_i|\vec{z}-\vec{z}_{0}|^{4}}\over
{[|\vec{x}-\vec{x}_{0\,i}|^2|\vec{z}-\vec{z}_{0}|^{2}+Q]^4}},\ \ \ 
H_3={Q\over {|\vec{z}-\vec{z}_{0}|^4}}.
\label{dinstharm}
\end{equation}
This is the solution constructed in Ref. \cite{kl}, where the 
radial coordinate $z=|\vec{z}-\vec{z}_{0}|$ in (\ref{dinstharm}) is 
the inverse of that in their solution.  
The worldvolume theory of the corresponding microscopic 
$D$-brane configuration is the multi-Yang-Mills instantons in 
the 4-dimensional $U(N_3)$ gauge theory with the instanton with 
the instanton number $N_{-1\,i}$ located at $\vec{x}=\vec{x}_{0\,i}$.  
Since the $D(-1)$-brane is located at the position where the 
$D3$-branes coincide, the size of the Yang-Mills instantons is 
infinite \cite{wit3}.

\subsection{Fundamental strings ending on $Dp$-branes}

There are $8-p$ overall transverse directions.  The fundamental 
strings have $p$ relative transverse directions and the $Dp$-branes 
have 1 relative transverse direction.  It is natural to let 
the solution to be delocalized along the relative transverse direction 
of the $Dp$-branes.  The spacetime metric has the following form:
\begin{eqnarray}
ds^2_{10}&=&-H^{-1}_FH^{-1/2}_pdt^2+H^{-1/2}_p(dx^2_1+\ldots+
dx^2_p)
\cr
& &+H^{-1}_FH^{1/2}_pdy^2+H^{1/2}_p(dz^2_1+\ldots+dz^2_{8-p}),
\label{fdp}
\end{eqnarray}
where the harmonic functions $H_F$ and $H_p$ for the fundamental 
string and $Dp$-branes satisfy the equations:
\begin{equation}
\partial^{2}_{\vec{z}}H_{F}+H_{p}\partial^{2}_{\vec{x}}H_{F}=0, 
\ \ \ 
\partial^2_{\vec{z}}H_p=0.
\label{fdplap}
\end{equation}
The harmonic functions are therefore given by
\begin{equation}
H_F=1+\sum_i{{Q_i}\over{[|\vec{x}-\vec{x}_{0\,i}|^2+
{{4Q}\over{(p-4)^2}}|\vec{z}-\vec{z}_{0}|^{p-4}]^{{p^2-6p+12}
\over{2(p-4)}}}},\ \ \ H_p={Q\over {|\vec{z}-\vec{z}_{0}|^{6-p}}}.
\label{fdpharm}
\end{equation}

Note that for $p<4$ the above solution becomes delocalized and 
for $p=4$ the solution becomes singular.  After $5-p$ overall transverse 
directions are delocalized, the harmonic functions take the following 
forms:
\begin{equation}
H_F=1+\sum_i{{Q_i}\over{[|\vec{x}-\vec{x}_{0\,i}|^2+
4Q|\vec{z}-\vec{z}_{0}|]^3}},\ \ \ 
H_p={Q\over {|\vec{z}-\vec{z}_{0}|}},
\label{fd4harm}
\end{equation}
representing fundamental strings localized except for the delocalized 
$5-p$ overall transverse directions.

For $p=3$, this supergravity solution is $S$-dual to the 
$D$-strings ending on $D3$-branes discussed in the previous 
section.  Once again, the ends of the fundamental strings 
are interpreted as (electrically) charged gauge bosons in 
the worldvolume gauge theory of the $D3$-branes.

\subsection{$Dp$-branes ending on $NS5$-branes}

There are 3 overall transverse directions.  The dimensions of the 
relative transverse spaces are $6-p$ for $Dp$-branes and 1 for 
$NS5$-branes.  It is natural to delocalize in the direction 
relatively transverse to the $NS5$-branes.  The spacetime metric 
for the $Dp$-branes ($p\leq 6$) ending on $NS5$-branes is 
given by
\begin{eqnarray}
ds^2_{10}&=&H^{-1/2}_p(-dt^2+dw^2_1+\ldots+dw^2_{p-1})+
H^{1/2}_p(dx^2_1+\ldots+dx^2_{6-p})
\cr
& &+H^{-1/2}_pH_{NS}dy^2+H^{1/2}_pH_{NS}(dz^2_1+dz^2_2+dz^2_3),
\label{dpns}
\end{eqnarray}
where the harmonic functions $H_p$ and $H_{NS}$ for the $Dp$-branes and 
$NS5$-branes satisfy the equations:
\begin{equation}
\partial^{2}_{\vec{z}}H_{p}+H_{NS}\partial^{2}_{\vec{x}}H_{p}=0, 
\ \ \ 
\partial^2_{\vec{z}}H_{NS}=0.
\label{dpnslap}
\end{equation}
The harmonic functions are therefore given by
\begin{equation}
H_p=1+\sum_i{{Q_i}\over{[|\vec{x}-\vec{x}_{0\,i}|^2+
4Q|\vec{z}-\vec{z}_{0}|]^{{8-p}\over 2}}},\ \ \ 
H_{NS}={Q\over {|\vec{z}-\vec{z}_{0}|}}.
\label{dpnsharm}
\end{equation}
The $p=6$ case is the supergravity solution for $NS5$-branes within 
$D6$-branes constructed in Ref. \cite{ity}.  

Since the $y$-direction, in which the $NS5$-branes stack, is 
delocalized, one can think of the $NS5$-branes as being periodically 
arrayed in the $y$-direction with the periodicity given by the 
circumference of the compactification circle.  
The scalars describing the fluctuations of the location 
$\vec{z}=\vec{z}_0$ of the $Dp$-branes in the perpendicular 
directions to the $NS5$-branes, along with the $y$-component 
of the $Dp$-worldvolume gauge field, form the hypermultiplet.  
The scalars corresponding to the locations $\vec{x}=\vec{x}_{0\,i}$ 
of the $Dp$-branes in the parallel directions of the $NS5$-branes, 
together with the $t$- and $\vec{x}$-components of the 
$Dp$-worldvolume gauge field, form the vector multiplet. Since 
the $Dp$-branes are finite in $y$, the worldvolume gauge theory 
is effectively $1+(p-1)$ dimensional with the Kaluza-Klein 
excitations on the $Dp$-branes invisible at large distances.  
For $p=3,4$, the corresponding worldvolume theory is $1+(p-1)$ 
dimensional $\prod^{N_5}_{i=1}U(N_4)$ supersymmetric gauge 
theory with $N_5$ (bifundamental) hypermultiplets transforming 
in the $(N_4,\bar{N}_4)$ of $U(N_4)\times U(N_4)$.  Here, each 
$U(N_4)$ gauge group is associated with $N_4$ $Dp$-branes that 
stretch between two adjacent $NS5$-branes.

\subsection{Fundamental strings with $NS5$-branes}

For fundamental strings parallel to $NS5$-branes, there are 4 overall 
transverse directions.  So, the delocalization along one of the overall 
transverse directions is required.  The fundamental strings have 4 
relative transverse directions and the $NS5$-branes have no relative 
transverse direction.  The metric has the following form:
\begin{equation}
ds^2_{10}=H^{-1}_F(-dt^2+dw^2)+dx^2_1+\cdots+dx^2_4
+H_{NS}(dz^2_1+\cdots+dz^2_4),
\label{fns}
\end{equation}
where the harmonic functions $H_F$ and $H_{NS}$ for the fundamental 
strings and the $NS5$-branes satisfy
\begin{equation}
\partial^{2}_{\vec{z}}H_{F}+H_{NS}\partial^{2}_{\vec{x}}H_{F}=0, 
\ \ \ 
\partial^2_{\vec{z}}H_{NS}=0.
\label{fnslap}
\end{equation}
The harmonic functions are therefore given by
\begin{equation}
H_F=1+\sum_i{{Q_i}\over{[|\vec{x}-\vec{x}_{0\,i}|^2+
4Q|\vec{z}-\vec{z}_{0}|]^3}},\ \ \ 
H_{NS}={Q\over {|\vec{z}-\vec{z}_{0}|}}.
\label{fnsharm}
\end{equation}
This solution is also constructed in Ref. \cite{ity}.

\subsection{Two $NS5$-branes intersecting over 3 dimensions}

The spacetime metric has the following form:
\begin{eqnarray}
ds^2_{10}&=&-dt^2+dw^2_1+dw_2^2+dw^2_3+H_1(dx^2_1+dx^2_2)
\cr
& &+H_2(dy^2_1+dy^2_2)+H_1H_2(dz^2_1+dz^2_2),
\label{nsns}
\end{eqnarray}
where the harmonic functions $H_1$ and $H_2$ for each 
$NS5$-branes satisfy:
\begin{equation}
\partial^2_{\vec{z}}H_1+H_2\partial^2_{\vec{x}}H_1=0, 
\ \ \ \  \partial^2_{\vec{z}}H_2=0.
\label{nsnslapl}
\end{equation}
The overall transverse space is 2-dimensional.  This means 
that the harmonic function $H_2$ in Eq. (\ref{harmtwo}) has 
to be logarithmic.  So, one cannot construct the localized 
solution by applying the method developed in this paper. 

\subsection{The KK monopole in the transverse space of $Dp$-brane 
with $p\leq 4$.}

The spacetime metric has the following form:
\begin{eqnarray}
ds^2_{10}&=&H^{-1/2}_p(-dt^2+dw^2_1+\ldots+dw^2_p)
+H^{1/2}_p[dx^2_1+\ldots+dx^2_{5-p}
\cr
& &+H_K(dz^2_1+dz^2_2+dz^2_3)+H^{-1}_K(dy+A_idz_i)^2], 
\label{kkdp}
\end{eqnarray}
where the harmonic functions $H_p$ and $H_{KK}$ for the $Dp$-branes 
and the KK monopoles and a 1-form potential ${\bf A}=(A_{i})$ satisfy
\begin{equation}
\partial^{2}_{\vec{z}}H_{p}+H_{KK}\partial^{2}_{\vec{x}}H_{p}=0, 
\ \ \ 
\partial^2_{\vec{z}}H_{KK}=0,\ \ \
\partial_{z_{i}}H_{K}=\epsilon_{ijk}\partial_{z_{j}}A_{k}.
\label{kkdplap}
\end{equation}
In the core region of the KK monopole or in the limit of large KK 
monopole charge, the harmonic functions are given by:
\begin{eqnarray}
H_{K}&=&{{Q_{KK}}\over{|\vec{z}-\vec{z}_{0}|}},\ \ \ \ 
{\bf A}=Q_{KK}\cos\theta d\phi,
\cr
H_{p}&=&1+\sum_i{{Q_{i}}\over{(|\vec{x}-\vec{x_{0\,i}}|^2
+4Q_{{KK}}|\vec{z}-\vec{z}_{0}|)^{{7-p}\over 2}}}.
\label{kkdpharm}
\end{eqnarray}

The corresponding worldvolume theory is the one with the flat 
transverse space replaced by an ALE space with $A_{N_{KK}-1}$ 
singularity.

\subsection{The KK monopole in the transverse space of the 
fundamental string}

The spacetime metric is given by
\begin{eqnarray}
ds^2_{10}&=&H^{-1}_F(-dt^2+dw^2)+dx^2_1+\ldots+dx^2_4
\cr
& &+H_K(dz^2_1+dz^2_2+dz^2_3)+H^{-1}_K(dy+A_idz_i)^2, 
\label{kkf}
\end{eqnarray}
where the harmonic functions $H_F$ and $H_{KK}$ for the fundamental 
strings and the KK monopole and a 1-form potential 
${\bf A}=(A_{i})$ satisfy
\begin{equation}
\partial^{2}_{\vec{z}}H_{F}+H_{KK}\partial^{2}_{\vec{x}}H_{F}=0, 
\ \ \ 
\partial^2_{\vec{z}}H_{KK}=0,\ \ \ 
\partial_{z_{i}}H_{K}=\epsilon_{ijk}\partial_{z_{j}}A_{k}.
\label{kkflap}
\end{equation}
In the core region of the KK monopole or in the limit of large KK 
monopole charge, the harmonic functions are given by:
\begin{eqnarray}
H_{K}&=&{{Q_{KK}}\over{|\vec{z}-\vec{z}_{0}|}},\ \ \ \ 
{\bf A}=Q_{KK}\cos\theta d\phi,
\cr
H_{F}&=&1+\sum_i{{Q_{i}}\over{(|\vec{x}-\vec{x_{0\,i}}|^2
+4Q_{{KK}}|\vec{z}-\vec{z}_{0}|)^3}}.
\label{kkfharm}
\end{eqnarray}

The corresponding worldvolume theory is described by a 
conformal field theory in the target manifold including 
an ALE space with $A_{N_{KK}-1}$ singularity. 

\subsection{The KK monopole in the transverse space of the 
$NS5$-brane}

The spacetime metric has the following form:
\begin{eqnarray}
ds^2_{10}=-dt^2+dw^2_1+\ldots+dw^2_5
+H_{NS}[H_K(dz^2_1+dz^2_2+dz^2_3)+H^{-1}_K(dy+A_idz_i)^2],
\label{kkn5}
\end{eqnarray}
where $H_{NS}$ and $H_K$ are the harmonic functions for 
the $NS5$-branes and the KK monopole and ${\bf A}=(A_{i})$ 
is a 1-form potential.  There are no relative transverse directions.  
So, there is no point in considering the localized intersecting 
configuration.

\subsection{The pp wave propagating in the background of the 
KK monopole}

The spacetime metric has the following form:
\begin{eqnarray}
ds^{2}_{10}&=&-dt^{2}+dw^{2}+(H_{W}-1)(dt-dw)^{2}+
dx^{2}_{1}+\ldots+dx^{2}_{4}
\cr
& &+H_{K}(dz^{2}_{1}+dz^{2}_{2}+dz^{2}_{3})+
H^{-1}_{K}(dy+A_{i}dz_{i})^{2}, 
\label{kkppten}
\end{eqnarray}
where the harmonic functions $H_{K}$ and $H_{W}$ for the 
KK monopole and the pp wave and a 1-form potential 
${\bf A}=(A_{i})$ satisfy the equations:
\begin{equation}
\partial^{2}_{\vec{z}}H_{K}=0,\ \ \ 
\partial_{z_{i}}H_{K}=\epsilon_{ijk}\partial_{z_{j}}A_{k},\ \ \ 
\partial^{2}_{\vec{z}}H_{W}+H_{K}\partial^{2}_{\vec{x}}H_{W}=0.
\label{kkpptenlap}
\end{equation}
In the core region of the KK monopole or in the limit of large KK 
monopole charge, the harmonic functions are therefore given by:
\begin{eqnarray}
H_{K}&=&{{Q_{KK}}\over{|\vec{z}-\vec{z}_{0}|}},\ \ \ \ 
{\bf A}=Q_{KK}\cos\theta d\phi,
\cr
H_{W}&=&1+{{Q_{W}}\over{(x^{2}+4Q_{{KK}}|\vec{z}-\vec{z}_{0}|)^3}}.
\label{kkpptenharm}
\end{eqnarray}

\subsection{Fundamental string with the pp wave propagating along 
its longitudinal direction}

The spacetime metric has the following form:
\begin{equation}
ds^{2}_{10}=H^{-1}_{F}[-dt^{2}+dw^{2}+
(H_{W}-1)(dt-dw)^{2}]+dz^{2}_{1}+\ldots+dz^{2}_{8},
\label{fpp}
\end{equation}
where $H_F$ and $H_{W}$ are respectively the harmonic functions for 
the fundamental strings and the pp wave.  Since there is no relative 
transverse directions, it is of no point to discuss the special 
localized solution.

\subsection{$NS5$-brane with the pp wave propagating along one 
of its longitudinal directions}

The spacetime metric has the following form:
\begin{eqnarray}
ds^{2}_{10}&=&-dt^{2}+dw^{2}+(H_{W}-1)(dt-dw)^{2}+dx^{2}_{1}
+\ldots+dx^{2}_{4}
\cr
& &\ \ \ \ \ +H_{NS}(dz^{2}_{1}+\ldots+dz^{2}_{4}),
\label{nspp}
\end{eqnarray}
with the harmonic functions $H_{NS}$ and $H_{W}$ for the $NS5$-brane 
and the pp wave satisfying the following equations:
\begin{equation}
\partial^{2}_{\vec{z}}H_{NS}=0,\ \ \ \  
\partial^{2}_{\vec{z}}H_{W}+H_{NS}\partial^{2}_{\vec{x}}H_{W}=0.
\label{nspplap}
\end{equation}
In the core region of the $NS5$-brane with one of its overall 
transverse directions delocalized, the harmonic functions are 
therefore given by:
\begin{equation}
H_{NS}={P \over {|\vec{z}-\vec{z}_{0}|}},\ \ \ \ 
H_{W}=1+{{Q_{W}}\over{(x^{2}+4P|\vec{z}-\vec{z}_{0}|)^3}}.
\label{nsppharm}
\end{equation}

\subsection{$Dp$-brane with the pp wave propagating along one 
of its longitudinal directions}

The spacetime metric has the following form:
\begin{eqnarray}
ds^{2}_{10}&=&H^{-1/2}_p[-dt^{2}+dw^{2}+
(H_{W}-1)(dt-dw)^{2}+dx^{2}_{1}+\ldots+dx^{2}_{p-1}]
\cr
& &\ \ \ \ \ +H^{1/2}_p(dz^{2}_{1}+\ldots+dz^{2}_{9-p}),
\label{dpp}
\end{eqnarray}
with the harmonic functions $H_p$ and $H_{W}$ for the $Dp$-brane 
and the pp wave satisfying the following equations:
\begin{equation}
\partial^{2}_{\vec{z}}H_{p}=0,\ \ \ \  
\partial^{2}_{\vec{z}}H_{W}+H_{p}\partial^{2}_{\vec{x}}H_{W}=0.
\label{dpplap}
\end{equation}
In the core region of the $Dp$-branes, the harmonic functions 
are therefore given by:
\begin{equation}
H_{p}={{Q}\over{|\vec{z}-\vec{z}_{0}|^{7-p}}},\ \ \ \ 
H_{W}=1+{{Q_{W}}\over{(x^{2}+{{4Q}\over{(p-5)^2}}
|\vec{z}-\vec{z}_{0}|^{p-5})^{{p^2-8p+19}\over{2(p-5)}}}}.
\label{dppharm}
\end{equation}
For the $p<5$ the pp wave becomes delocalized and for $p=5$ the above 
solution becomes singular.  After $6-p$ of the overall transverse 
directions are delocalized, the harmonic functions take the following 
forms:
\begin{equation}
H_{p}={{Q}\over{|\vec{z}-\vec{z}_{0}|}},\ \ \ \ 
H_{W}=1+{{Q_{W}}\over{(x^{2}+4Q|\vec{z}-\vec{z}_{0}|)^3}}.
\label{d4ppharm}
\end{equation}
The $p=6$ case is the supergravity solution for the pp wave localized 
within $D6$-branes constructed in Ref. \cite{ity}.

\section*{Acknowledgements}

I would like to thank Y. Oz for discussions. 


\end{document}